\documentclass[twoside,reqno]{bjp}
\usepackage{epsfig,cite}
\usepackage{graphics}
\usepackage{amssymb,amsmath}
\usepackage{times}
\begin{document}


% use the thanks command within \title, \author or \address for footnotes;
% \title{Title} or  \title{Title\thanks{...}}
% \author{Name1}{aff.label1}, \coauthor{Name2}{aff.label2},  \coauthor{Name3}{aff.label3}
% \address{Address1}{aff.label1}
% \address{Address2}{aff.label2}
% \address{Address3}{aff.label3}
%\runningheads{NAMES OF AUTHORS IN CAPITALS}{SHORT TITLE IN CAPITALS}

\title{Neutron Transfer Reactions for Deformed Nuclei Using Sturmian Basis States
%\thanks{This work was partly performed under the auspices of the U. S. Department of Energy by the University of California, Lawrence Livermore National Laboratory under contract No. DE-AC52-07NA27344. The authors are very grateful to Peter D. Kunz and Frank S. Dietrich for essential input and guidance in performing some of the calculations and discussions in the early stages of this research. } -- JE: this is already at the end of the paper 
}

\runningheads{V.~G.~GUEORGUIEV and J.~E.~ESCHER}
{NEUTRON TRANSFER REACTIONS FOR DEFORMED NUCLEI ...}

\begin{start}

\author{V.~G.~Gueorguiev}{1,2} and \coauthor{J.~E.~Escher}{3} %,\coauthor{P.~D.~Kunz}{4},\coauthor{F.~S.~Dietrich}{3}

\address{Ronin Institute, NJ, USA}{1}
\address{Institute for Advanced Physical Studies, Sofia, Bulgaria}{2}
\address{Lawrence Livermore National Laboratory, Livermore, California, USA}{3}
%\address{Dept. of Physics and Astrophysics, University of Colorado, Boulder, USA}{4}

\received{31 October 2017}

\begin{Abstract}
We study the spin-parity distribution P(J$^{\pi}$,E) of $^{156}$Gd excited states above the neutron separation energy $S_{n}=8.536$ MeV \cite{156Gd_list_of_levels} that are expected to be populated via the 1-step neutron pickup reaction $^{157}$Gd($^{3}$He,$^{4}$He)$^{156}$Gd. In analogy with the rotor plus particle model \cite{Bohr&Mottelson-II}, we view excited states in $^{156}$Gd as rotational excitations built on intrinsic states consisting of a neutron hole in the $^{157}$Gd core; that is, a neutron removal from a deformed Woods-Saxon type single-particle state \cite{Woods-Saxon:1954} in $^{157}$Gd. The particle-core interaction usually dominated by a Coriolis coupling are accounted via first order perturbation theory \cite{VGGueorguiev:07062002}. The reaction cross section to each excited state in $^{156}$Gd is calculated as coherent contribution using a  standard reaction code \cite{CHUCK} based on spherical basis states. The spectroscopic factor associated with each state is the expansion coefficient of the deformed neutron state in a spherical Sturmian basis along with the spherical form factors \cite{VGGueorguiev:07062002}. The total cross section, as a function of the excitation energy, is generated using Lorentzian smearing distribution function. Our calculations show that, within the assumptions and computational modeling, the reaction $^{3}$He+$^{157}$Gd $\rightarrow$ $^{4}$He+$^{156}$Gd$^{\star}$ has a smooth formation probability P(J$^{\pi}$,E) within the energy range relevant to the desired reaction $^{155}$Gd+n $\rightarrow$ $^{156}$Gd$^{\star}$. The formation probability P(J$^{\pi}$,E) resembles a Gaussian distribution with centroids and widths that differ for positive and negative parity states.
\end{Abstract}

\end{start}

%Relevant topics by PACS number:\newline
%\noindent
%02.60.Dc Numerical linear algebra\newline
%21.10.Jx Spectroscopic factors\newline
%21.10.Re Collective levels\newline
%21.10.Pc Single-particle levels\newline
%21.60.Cs Shell model\newline
%24.50.+g Direct reactions\newline
%24.10.Ht Optical and diffraction models\newline
%24.10.Eq Coupled-channel and distorted wave models\newline
%25.55.-e 3H-, 3He-, and 4He-induced reactions\newline
%25.55.Ci Elastic and inelastic scattering\newline
%25.55.Hp Transfer reactions\newline
%25.70.De Coulomb excitation\newline
%25.70.Gh Compound nucleus\newline
%27.70.+q 150 $\leq $ A $\leq $ 189\newline
%\newline
%Subject keywords: 
%Single-particle levels for axially-deformed Woods-Saxon potential;
%Spectroscopic factors within the particle-plus-rotor model; 
%Direct reaction cross sections for strongly deformed nuclei;

\section[]{Introduction and Motivation}
% JE: shortening this to the most important points to save space
%Most light nuclei up to C and O are produced via nuclear reactions within the stars. Heavier nuclei near the valley of stability are produced via slow neutron capture process. To account for the observed abundance of the elements, however, one needs to also consider other processes such as proton capture process and rapid neutron capture process (r process). The r process is essential in the production of the heaviest neutron-rich elements since this process generates nuclei far from the valley of stability that decay back towards the valley. Successively heavier neutron-rich nuclei are produced through violent processes such as supernova explosions. Such nuclei usually decay quickly. As a consequence, the understanding of the r-process contribution to the 

\begin{figure}[tbh]
\centerline {\includegraphics[width=10cm]{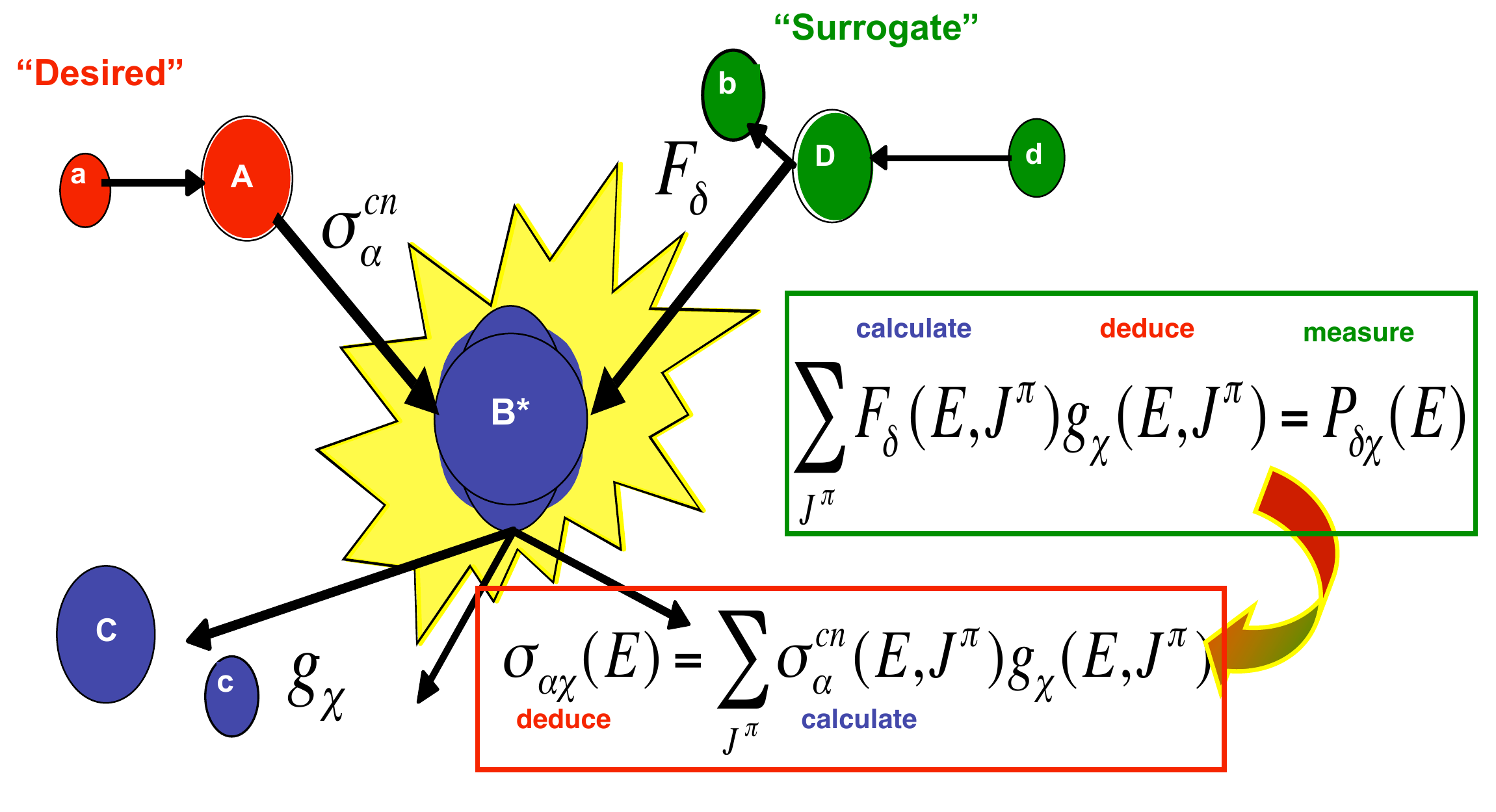}}
\caption{Surrogate idea. When the desired reaction a+A $\rightarrow B^{\star}\rightarrow$ C+c involves an unstable target A, the cross section may be deduced using theoretical modeling and experimental data from a surrogate reaction d+D $\rightarrow b+B^{\star} \rightarrow$ C+c on a stable target D. Both reactions proceed via the same compound system B$^{\star}$.}
\label{TheSurrogateMethod}
\end{figure}

Understanding the production of the heavy elements is one of the most important challenges for nuclear astrophysics~\cite{Arcones:17}.
Multiple nucleosynthesis processes play a role and unravelling their respective contributions to the 
observed abundances of the elements requires knowledge of the neutron-induced reactions on unstable nuclei. Unfortunately,  measuring these cross sections in a laboratory environment is a very difficult, if not impossible, task because of the technical and practical problems associated with the use of an unstable nuclei. There have been various proposals for circumventing the problems associated with the use of unstable nuclei and yet to gain information about the desired nuclear reaction. One such approach is the Surrogate Method~\cite{Escher:12rmp} shown schematically in Fig. \ref{TheSurrogateMethod}.

In this paper we study the reaction $^{3}$He+$^{157}$Gd $\rightarrow$ $^{4}$He+$^{156}$Gd$^{\star}$ and model the formation probability of various excited states of the $^{156}$Gd system within a direct reaction framework. This reaction is a surrogate for the neutron capture  reaction $^{155}$Gd+n $\rightarrow$ $^{156}$Gd$^{\star}$. To study excitation energies $E_{ex}$ of $^{156}$Gd that correspond to low-energy neutron absorption by $^{155}$Gd, we consider $E_{ex} >$  $S_{n}=8.536$ MeV, where $S_{n}$ is the neutron separation energy for $^{156}$Gd \cite{156Gd_list_of_levels}.
%JE: shortening text
%by energy balance one has: $E_{ex}=M_{^{155}Gd}+M_{n}+E_{n}-M_{^{156}Gd}=S_{n}+E_{n}$ where M is the corresponding rest mass and $S_{n}=8.536$ MeV is the neutron separation energy for $^{156}$Gd \cite{156Gd_list_of_levels}. Thus, we will focus on $^{156}$Gd excited states with $E_{ex}$ >  8.5 MeV.

\section{Mathematical Framework}
Atomic nuclei exhibit a multitude of spectral phenomena, such as
%. Among the simplest and relatively well understood properties are the 
rotational and vibrational spectra, as well as the single-particle shell structure that explains the observed magic numbers \cite{Bohr&Mottelson-I,Bohr&Mottelson-II}. Usually, rotational  states of deformed even-even nuclei are modeled using simple rotational functions \cite{Bohr&Mottelson-II}. An extra particle is then assumed to occupy a valence single-particle state within a deformed mean-field potential. This allows one to treat the odd-even deformed nuclei within a model consisting of a rotor plus a single particle. Within this framework one can calculate cross sections for particle transfer using transition amplitudes between collective rotational states and states of a particle plus a rotor.

\subsection{Energy levels of an axially-deformed rotor plus a particle system}\label{SoADN}
Within the Bohr-Mottelson rotor model, an intrinsic state $\Phi _{K}$ of a deformed system, with axial symmetry, gives rise to a rotational band $E(J)=E_{K}+\frac{\hbar^{2}}{2\cal{I}}(J(J+1)-K(K+1))$. Here $K$ is the angular momentum projection onto the symmetry axis and $\cal{I}$ is the moment of inertia of the system, which for simplicity is assumed to be independent of $K$ \cite{Bohr&Mottelson-II}. In the zero order approximation, where we neglect possible particle-core couplings, such as Coriolis coupling, the combined system of a particle/hole plus a core has an  intrinsic state $\Phi _{\Omega}$ that can be viewed as a direct product of the  intrinsic state of the core $\Phi^{cor}_{K}$ and the single-particle/hole state  $\psi_{\nu}$ with an angular momenta projection $\nu$:
$$\Phi _{\Omega}=\psi_{\nu}\Phi^{cor}_{K},~E_{\Omega}=E_{K}+\epsilon_{\nu},$$
where  $\Omega=K+\nu$ due to axial symmetry. Therefore, for the particle-plus-core system we have:
$E(J)=E_{\Omega}+\frac{\hbar^{2}}{2\cal{I}}(J(J+1)-\Omega(\Omega+1)).$ 

As long as the states of interest are built from the same initial intrinsic core state, we can make perturbation adjustments to the energies to incorporate pairing and Coriolis coupling effects. Coriolis coupling $H_{C}=-\frac{\hbar^{2}}{2\cal{I}}(I_{+}j_{-}+I_{-}j_{+})$ is known to be an important interaction, where $I_{\pm}$ and $j_{\pm}$ are the ladder spin operators for the core and single particle \cite{Kanestrom&Tjom:1969,Joshi&Sood:1974,Peng&Maher:1976,granja:034316}. It is known that the first-order perturbation to the energy is non-zero for $\Omega=\frac{1}{2}$ bands only and higher order terms are needed to obtain the energy shifts in $\Omega>\frac{1}{2}$ bands \cite{Kanestrom&Tjom:1969,Joshi&Sood:1974}. We consider only one intrinsic state of the core (the ground state of the target) coupled to various neutron holes in the core. This, therefore, eliminates any Coriolis band mixing from our model.  
%JE: not needed, as this is not done here:
%However, there is an important particle/hole-core coupling that can be estimate by using first-order perturbation approximation which gives us the  energy splitting between the states $\Omega=|K-\nu|$ and $\Omega=K+\nu$. To estimate the splitting of these two states, due to rotation, we look at the expectation value of the rotational energy of the particle-plus-core system in the intrinsic frame where two of the terms would represent the rotational energy of the core  and the particle and the energy splitting would then be: $\Delta E(\Omega_{\pm}=|K\pm\nu|)=\pm c\frac{\hbar^{2}}{2\cal{I}}K\nu$. Here $c$ is a coefficient that relates the moment of inertia around the symmetry axis $z$ to the moment of inertia around $x$ and $y$ axes.  One could estimate its value from the mathematical expressions for the  moment of inertia of a rigid ellipsoid as given in the Appendix section \cite{VGGueorguiev:07062002}. Here, however, we prefer 
We use a phenomenological adjustment that reproduces the experimentally observed splitting between $\Omega =0$ and $\Omega=3$ bands in $^{156}$Gd by using the experimental moment of inertia $\cal{I}$ needed to reproduce the $\Omega =0$ ground state band.
We have already included particle-core effects due to the Coriolis coupling, but will neglect the pairing effects. 
%JE: not needed, as this is not done here:
%The usual pairing interaction consists of a pair creation and annihilation operators - its main effect is on the occupation of the single-particle states near the Fermi energy that represents the location of the last occupied state and next unoccupied state. Thus, the occupation number $n_{\nu}$ of states that are sufficiently far from the Fermi energy $\mu$ are practically not affected by the paring interaction. We are looking at excited states in $^{156}$Gd with $E_{ex} >  8.5$ MeV that correspond to neutrons that are at least 8.7 MeV below the Fermi level ($\mu=-6.169$ MeV) and thus $\epsilon_{\nu}< -14.9$ MeV. By applying the BCS theory one can see that the occupation of the single particle levels with energies $\epsilon_{\nu}<-10$ MeV are practically unaffected by the pairing interaction \cite{VGGueorguiev:07062002}. 
This is justified as we are considering highly excited states in $^{156}$Gd with $E_{ex} >  8.5$ MeV.

\subsection{Description of the single-particle states in axially-deformed nuclei}
We calculate single-particle states using Woods-Saxon mean-field potential \cite{Woods-Saxon:1954}: 
%Since the nucleon density in nuclei is constant, except near the nuclear surface, one expects that the general form of the mean-field potential, which defines the single-particle states, is of Woods-Saxon type \cite{Woods-Saxon:1954}: 
$$V(r,R)=V_{0}/{(1+\exp \left({(r-R)/a}\right))}$$
For spherical nuclei $R$ is constant and represents the position of the nuclear surface while $a$ is a measure of the diffuseness of the potential near the surface. 
%JE: shorten
%For deformed nuclei, however, $R$ depends on the surface point of interest and is often parametrized using spherical harmonics $Y_{\lambda,m}\left( \theta ,\phi \right)$. Furthermore, for axially-deformed nuclei $R$ does not depend on the angle $\phi$ due to the axial symmetry. The monopole $\lambda=1$ term is absent due to center-of-mass considerations and odd $\lambda$ terms are usually absent due to parity considerations; for 
We consider axially-deformed nuclei and assume quadrupole and hexapole deformation only: 
$$R(\theta ,\phi )=R_{0}\left( 1+\beta _{2}Y_{20}\left( \theta ,\phi \right)+\beta _{4}Y_{40}\left( \theta ,\phi \right) \right)$$
For small deformations ($\beta \lesssim 0.3$) one can expand the Woods-Saxon potential in Taylor series. While in many cases it seems sufficient to consider only the first-order terms in the expansion \cite{Hird&Huang:CPC1975}, some single-particle states are sensitive to small values of beta $\beta_{2}\lesssim 0.3$. Proper treatment of such states needs careful considerations~\cite{VGGueorguiev:07062002}. If $\beta $ is sufficiently small, so that the Taylor expansion converges, one can re-express $V(r,R)$ in terms of spherical harmonics \cite{Glendenning:2004}. An alternative way is to solve numerically the Schr\"{o}dinger equation for the deformed Woods-Saxon potential \cite{Dudek&Nazarewicz}. When comparing the single-particle energies calculated numerically to the one calculated using only first-order Taylor expansion approximation, one finds that for rare-earth nuclei (nuclei near Gd) the ``Nilsson diagrams'' agree for $\beta _{2}\lesssim 0.1$ but start to deviate at larger deformations; in particular, there is a substantial deviation for $m_{j}=j$ states \cite{Glendenning:2004}. 

For our study the neutron bound states in $^{157}Gd$ were calculated with the WSBETA code \cite{Dudek&Nazarewicz} using Woods-Saxon parameters from Ref. \cite{Hird&Huang:CJP1975}, but $\beta_{4}=0$ so that the 47$^{th}$ neutron state ($\epsilon_{47}=-6.361$) is near the experimental neutron separation energy $S_{n}=6.3598$ MeV \cite{157Gd_list_of_levels}, $V_{0}$=-45.1776 MeV, $r_{0}$=1.25~fm, $a_{0}$=0.65~fm, $V_{ls}$=19.2015 MeV, $\beta_{2}=0.29$ and $\beta_{4}=0$ \cite{VGGueorguiev:07062002}.

\section{Reaction Cross Sections for Deformed Nuclei}

We employ the zero-range Distorted Wave Born Approximation (DWBA) to calculate the 1-neutron pickup reaction~\cite{Satchler:1958,DWUCK, CHUCK}.  To carry out our calculations, we need optical potentials, single-nucleon wave functions, and spectroscopic factors.
To calculate the distorted waves, we have used~\cite{VGGueorguiev:07062002} an Optical Model Potential (OMP) of Wood-Saxon type, with parameters from the Reference Input Parameter Library (RIPL-2) \cite{RIPL-2,Becchetti&Greenleesg:1969,Avrigeanu&Hodgson:1994}.

For transfer reactions, which result in low-energy excitations, the correct asymptotic tail of the wave function, which is related to the neutron separation energy, is very important in neutron pickup since transfer reactions are sensitive to the nuclear surface.
%JE: not a correct statement: one expects that the last (outer most) nucleon is being transferred. 
Using wave functions with the desired binding energy produced by adjusting the depth of the Woods-Saxon binding potential is one of the simplest and usually very successful method in calculating the reaction cross sections. In this  approach one keeps the geometric factors of the binding potential fixed from systematics, but changes the depth of the potential until a state $\psi$ with the desired binding energy is found. 

\subsubsection{Sturmian method for reaction form-factors and spectroscopic factors}

%JE: not necessary:
%The separation energy matching method seems very ad hoc. Despite its simplicity of using only one spherical wave function, conceptual and actual problems has been a motivation for more accurate description of the relevant form factor by inclusion of more than one basis state. The spherical harmonic oscillator (SHO) is a traditional basis to expand a wave function. However, the tail in a finite SHO basis expansion is always wrong. 
%An alternative basis for expanding a deformed state is to use the set of bound states of the spherical part of the interaction \cite{Hird&Huang:CJP1975}. However, this expansion does not guarantee the correct tail either. An expansion would provide the correct tail if one of the basis states is a bound state with the same energy as the state that is being expanded and all the other non-zero components correspond to basis states with a deeper binding energy. 

We employ a Sturmian basis to determine single-particle wave functions.  This approach has advantages over alternative methods.
In a Sturmian basis all the basis states have the same tail as the original state that is being expanded in this basis. In order to maintain the correct asymptotic tail one has to find different wave functions and potential strengths that result in the same energy and thus the same wave function tail.  The Sturmian basis method has been utilized before in transfer reactions to highly excited states in deformed nuclei \cite{Andersen&Back:1970}. The method in \cite{Andersen&Back:1970} relies essentially on expressing the deformed potential as linear in $\beta_{2}$ which may not be sufficient in the case of strong deformation.

% JE: Save the discussion for a long paper.
%Another, more successful and physically more appealing approach is to consider coupled-channel (CC) calculations \cite{Rost:1967}. Unfortunately, this code  in its current version has only $\beta_{2}$ deformation. In one of the next section we will discuss briefly our comparison to this code and possible further developments. 

To illustrate the role of various parameters involved, we now look at the incoherent DWBA reaction cross section for a particle transfer from a deformed single-particle state $\psi_{\nu}$ that can be expressed in terms of transfer cross sections on spherical single-particle states $\phi_{nlj}$ \cite{Satchler:1958,Andersen&Back:1970}: 

\qquad\qquad\qquad $d\sigma(J_{i}K_{i} \rightarrow J_{f}K_{f};\nu)=\sum_{lj}\sum_{n}(a_{\nu}v_{\nu}c_{\nu}^{nlj})^{2}d\sigma^{DW}_{nlj}$

Here $\sigma^{DW}_{nlj}$ are the DWBA cross section for pickup from a spherical state $\phi_{nlj}$, $c_{\nu}^{nlj}$ are the expansion coefficients of the  state $\psi_{\nu}$ in spherical basis states $\phi_{nlj}$, $v_{\nu}$ represents BCS occupation number ($v_{\nu}^{2}=n_{\nu}/2$) of the state $\nu$, and $a_{\nu}$ is the Coriolis band mixing amplitude. The spectroscopic factor $S_{lj}$ is often used as shorthand notation for the term $(a_{\nu}v_{\nu}c_{\nu}^{nlj})^{2}$ above. In our calculations, we actually consider the generally more appropriate coherent cross section by using super-position of basis states $\phi_{nlj}$ with amplitudes $a_{\nu}v_{\nu}c_{\nu}^{nlj}$.

In our study, the individual cross sections, for neutron transfer from a deformed state $\psi_{\nu}$ that results in a final state with $J^{\pi}$ and energy $E$, are calculated as coherent cross sections with the code CHUCK3 \cite{CHUCK} using the amplitudes $c^{\nu}_{nlj}$ times a 
Clebsch-Gordan coefficient and other appropriate factors  \cite{Andersen&Back:1970}:

$\sqrt{{(1+\delta_{0,K_{i}K_{f}})}/{(2j+1)}}\times D_{0}\times c^{\nu}_{nlj}\times(J_{f}K_{f}| j m, J_{i},K_{i}),$
where $D_{0}$  is the strength of the zero range transfer potential ($D_{0}\delta (x)$, $D_{0}^{2}$=18 \cite{DWUCK}). 
The resulting reaction cross-sections $\sigma_{\lambda}(\epsilon_{\nu},J^{\pi})$ is defined for each single-particle state $\epsilon_{\nu}$ . 
In what follows the index $\lambda$ will denote the pair of labels $(\epsilon_{\nu},J^{\pi})$.

\section{Neutron Transfer Reaction Results}
Here we present the results for the transfer reaction $^{157}${Gd}($^{3}${He},$^{4}${He})$^{156}${Gd}.   
After obtaining the individual cross-sections $\sigma_{\lambda}(\epsilon_{\nu},J^{\pi})$ within the Sturmian method one can consider the total cross-section $\sigma(E)$ using a smeared function, which takes into account damping effects that are not explicit in our model. Then, the compound-formation probability distributions $P(J^{\pi},E)$ can also be computed.

The transfer cross sections $\sigma_{\lambda}(\epsilon_{\nu},J^{\pi})$ are for one-nucleon removal from a deformed single-particle state $\psi_{\epsilon_{\nu}}$ from  the $^{157}$Gd system. The $^{157}$Gd system is the core in a $K=3/2^{-}$ state. We treat the final states of the $^{156}$Gd system as rotational states built on the intrinsic state $\Omega$ where $\Omega$ refers to neutron holes in the core $|A=156,{\Omega}>=\psi^{\dagger}_{\nu} ~|{A=157},K={3/2}^{-}>$.

In order to calculate the cross section for a one-nucleon transfer reaction to an excited state in $^{156}$Gd, we first consider an intrinsic state in $^{156}$Gd as a hole in the $^{157}$Gd core; then we construct rotational states built on that intrinsic state. 
%JE: rephrase
%Simple geometric estimate assuming a 60 MeV projectile with impact parameter twice the classical radius of $^{157}$Gd gives maximum angular momentum transfer of 4. 
We use simple geometric arguments to estimate the maximum angular momentum transfer to be 4; future work may relax this constraint.
Therefore,  one can consider a total of 12 members of the related $\Omega=|K\pm\nu|$ rotational bands to include all possible excitations with $J<8$. Thus, the relevant states are:
$|^{156}Gd,{\Omega=|K\pm\nu|}>=\psi^{\dagger}_{\pm\nu} |^{157}Gd,K={3/2}^{-}>$.
$$E(J^{\pi};\Omega=|K\pm\nu|)=\epsilon_{0}-\epsilon_{\nu}+\frac{\hbar^{2}}{2\cal{I}}(J(J+1)+\delta_{\pm})$$
Here $\epsilon_{0}$ is used to set the ground state energy of $^{156}Gd$ to zero, and $\delta_{\pm}$ is the energy shift of the $\Omega=K+\nu$ state relative to the $\Omega=|K-\nu|$ state. Thus if we set $\delta_{-}=0$ for $\Omega=|K-\nu|$ then $\delta_{+}=2cK\nu$ for $\Omega=K+\nu$ states. We always keep $\Omega,\nu$, and $K$ positive when we use them as labels for the states.

By fitting the first four excited states of the $\Omega=0^{+}$ ground state band, the moment of inertia is fixed to be ${\hbar^{2}/2\cal{I}}=13.59$ KeV for the rotational bands in $^{156}$Gd. The $c=17.741$ coefficient for the Coriolis energy shift of the related $\Omega=3^{+}$ band was chosen to reproduce the excitation energy of the $3_{1}^{+}$ state. The value of $c$ is about 1.65 times the ratio of the moments of inertia $\cal{I_{||}/\cal{I_{\perp}}}$ for rigid ellipsoid with $\beta=0.29$. This way were are within few keV of the experimental values for the $0^+$ ground state band and the $3^+$ band \cite{VGGueorguiev:07062002}.

%JE: not needed
%There is a 2 MeV gap between the single-particle states with energy -16.81 MeV and -14.66 MeV that results in absence of positive-parity states in the 10 MeV excitation energy region. This will produce a parity asymmetry in the $P(J^{\pi},E)$ distribution in the energy region of interest for the surrogate method \cite{VGGueorguiev:07062002}.

%HERENOW
\subsection{Neutron pickup cross sections}

%JE: Cut repetitive text to shorten paper:
%In order to compute the transfer cross section one needs the single-particle wave function and the corresponding spectroscopic factor. We already discussed our main approximations to the spectroscopic factor: we neglect the Coriolis band mixing ($a_{\nu}=1$), the pairing effects ($v_{\nu}=1$), and consider single-particle wave functions that are expressed in a Sturmian basis. The idea is to maintain the same exponential tail for each spherical basis state as the tail of the deformed state that we are interested in.  We have to make an expansion in a spherical basis in order to be able to use existing reaction codes. 

To be able to use existing reaction codes, we have to expand the deformed single-particle states in a spherical basis.
Here we consider the Sturmian approach by using Sturmian spherical basis states. We calculate the spectra of a deformed Woods-Saxon potential using standard bound-states technique and employ the code WSBETA \cite{Dudek&Nazarewicz}. Then for each state $\psi_{\nu}$ with energy $\epsilon$ we find all the  Sturmian spherical basis states (zero deformation) $\phi_{\epsilon n l j}$ with $nlj$ labels as for a spherical harmonic oscillator up to the $N_{max}$ oscillator shell. These basis states are constructed with the reaction code DWUCK4 \cite{DWUCK}. For a fixed $\epsilon$ and $nlj$ labels the code finds a scaling factor for the original spherical potential such that $\phi_{\epsilon n l j}$ is a bound state of this new potential. This scaling factor is then used to recompute the state $\phi_{\epsilon n l j}$ within the WSBETA code in the same basis where the deformed state $\psi_{\nu}$ has been computed. 
The expansion amplitudes $c_{\nu}^{nlj}$ are then calculated~\cite{VGGueorguiev:07062002} and passed to the reaction code CHUCK3 \cite{CHUCK}, which has the ability to add the $c_{\nu}^{nlj}$ amplitudes coherently.
%JE: shorten
%to compute the cross section for pickup from the deformed state $\psi_{\nu}$ using the spherical basis states $\phi_{\epsilon n l j}$. 
%Since the $c_{\nu}^{nlj}$ amplitudes have to be added coherently, we have used the coupled-channel code CHUCK3 \cite{CHUCK}.

%JE: I moved that up in the paper, to not interrupt the flow of the text here.
%Calculations with distorted waves within the DWBA use an Optical Model Potential (OMP) of Wood-Saxon type.  We have used~\cite{VGGueorguiev:07062002} parameters from the Reference Input Parameter Library (RIPL-2) \cite{RIPL-2,Becchetti&Greenleesg:1969,Avrigeanu&Hodgson:1994}.

\begin{figure}[htb]
\centerline{\includegraphics[width=10cm]{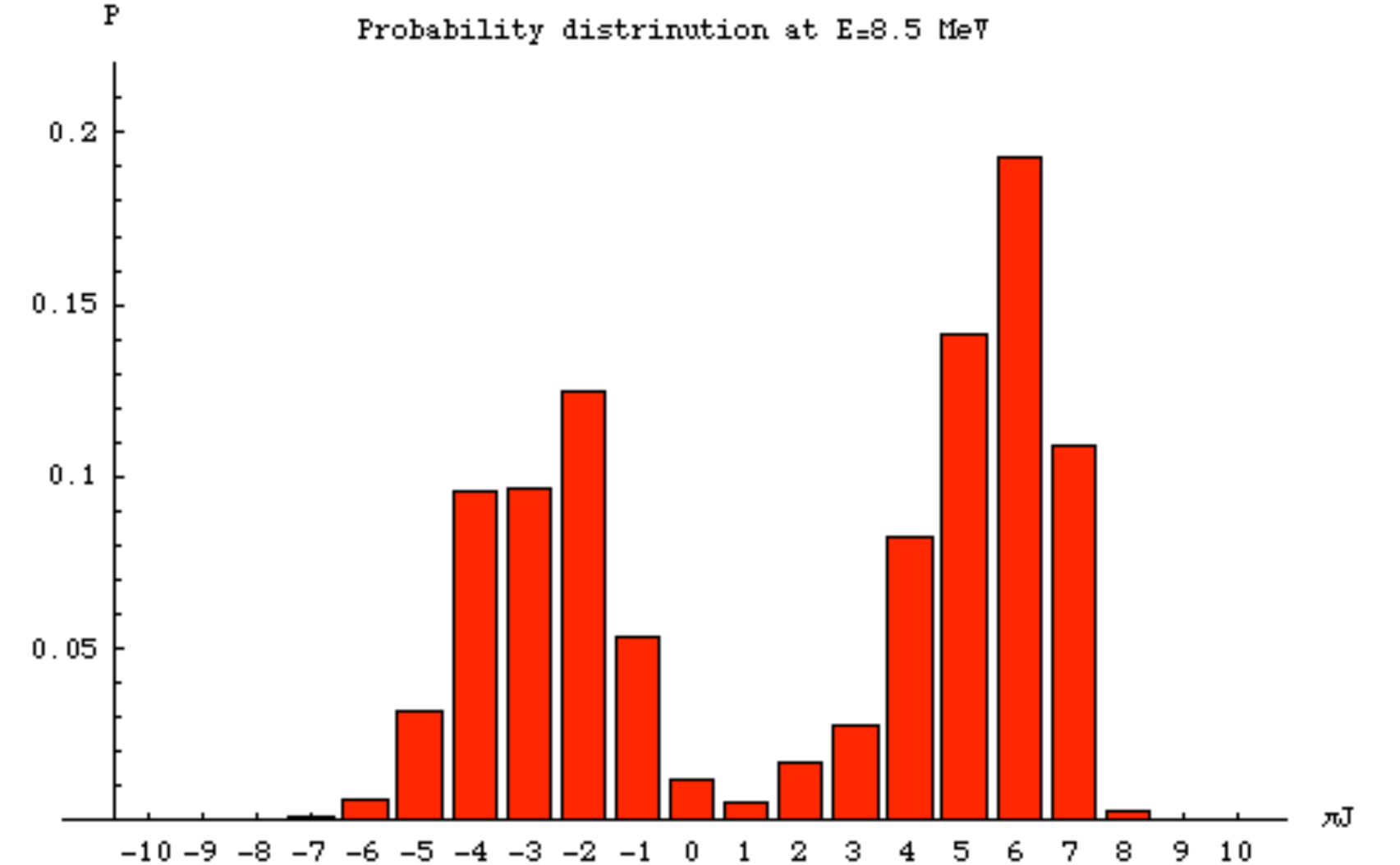}}
\centerline{\includegraphics[width=10cm]{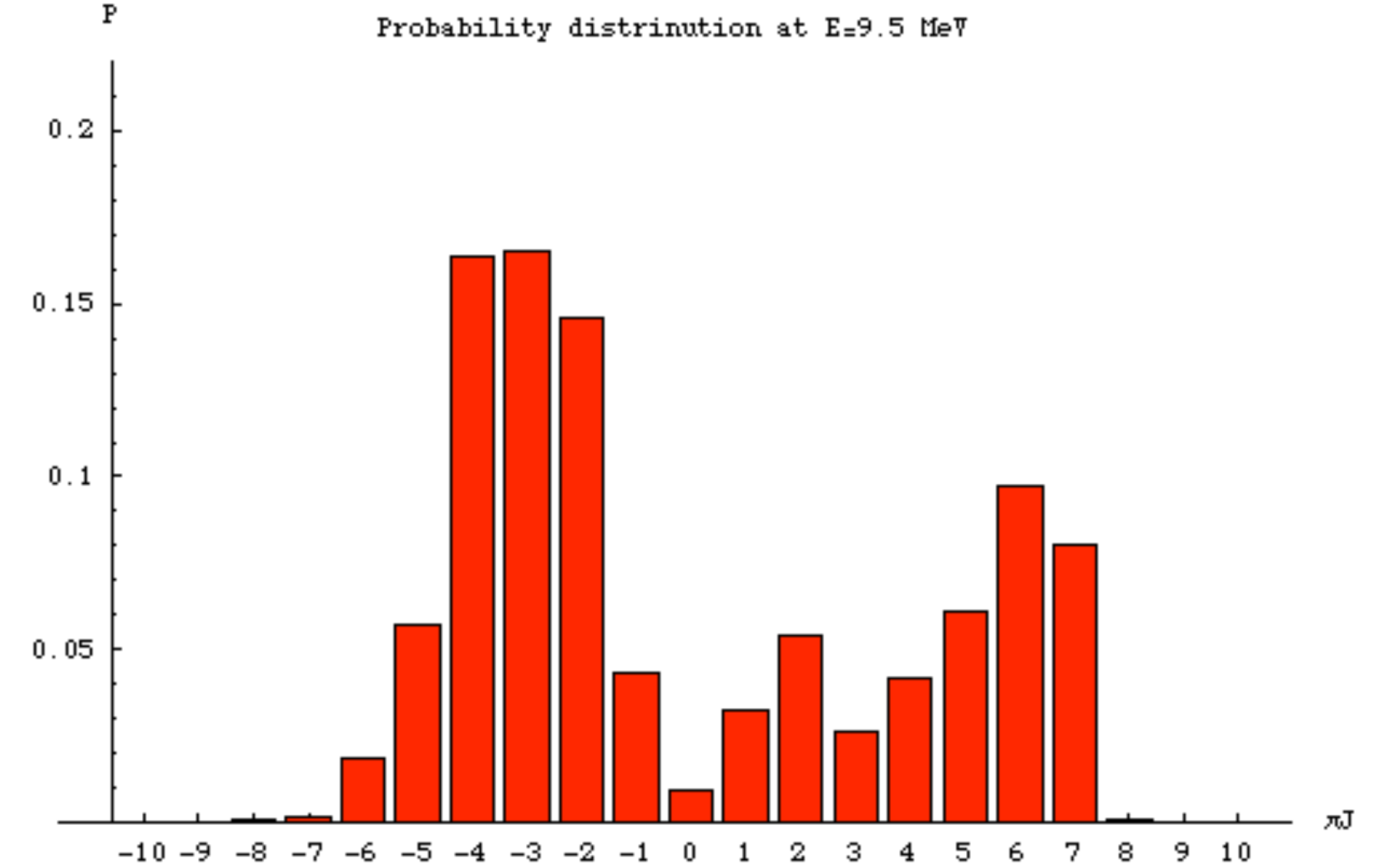}}
\caption{$P(J^{\pi};E)$ distributions for energies near the neutron separation energy 
in $^{156}$Gd. Top graph for $E=8.5$ MeV and bottom $E=9.5$ MeV (using $\Gamma=0.01+0.01E$ for the smearing function). 
The sign of the horizontal coordinate corresponds to the parity $\pi$ and its magnitude gives $J$.}
\label{3Jpi_dist}
\end{figure}

The cross sections that one can calculate within the presented framework correspond to sharp final states. In reality there are widths associated with the single-particle states as well as with the final states. In order to produce a smooth total cross section as a function of the excitation energy of the $^{156}$Gd system we consider a smearing distribution function of Lorentzian type: 
$$\rho_{\nu}(E)=\frac{1}{2\pi}\frac{4\Gamma}{4(E-E_{\nu})^{2}+\Gamma^{2}},$$ 
with $\Gamma=a+b E$ and define a smooth $\sigma(E)$ cross-section \cite{Andersen&Back:1970}:
$$\sigma(E)=\sum_{\lambda}\rho_{\lambda}(E)\sigma_{\lambda}$$
The smeared cross sections, introduced above, can be used to determine a smooth probability to excite a state with quantum numbers $J^{\pi}$:
$$P(J^{\pi};E)=\frac{1}{\sigma(E)}\sum_{\lambda}\delta_{J,J_{\lambda}}\delta_{\pi,\pi_{\lambda}}\rho_{\lambda}(E)\sigma_{\lambda}$$
In Fig. \ref{3Jpi_dist} we show the $P(J^{\pi};E)$ distributions that are of interest to the surrogate method through direct neutron pickup via 42 MeV $^{3}$He
on $^{157}$Gd target. 
We find that all final spins from $J = 0$  to $J = 8$ are populated. The calculated formation probability P(J$^{\pi}$,E) resembles Gaussian distributions with magnitudes, centroids, and width that are different for positive and negative parity states. The parity asymmetry in the $P(J^{\pi},E)$ distribution can be tracked back to the neutron single-particle states and their energies. There is a 2 MeV gap between the single-particle states with energy -16.81 MeV and -14.66 MeV that results in absence of positive-parity states in the 10 MeV excitation energy region \cite{VGGueorguiev:07062002}. The results are to be considered a first estimate for the spin-parity distribution. A more comprehensive treatment would need to relax the geometric estimate of an upper limit for the angular momentum transfer.  Also, recent work~\cite{Escher:16a} has demonstrated that two-step reaction mechanisms can play an important role in transfer reactions that populate highly-excited states (above a few MeV).
%JE: dead channels is confusing and would require too much detailed explanation. I simplified this. Also, I needed to say something about the limitations of the current approach.
%These graphs show that at least in principle there are no dead channels due to zero  $P(J^{\pi})$ and thus any channel that goes through $J < 8$ should be able to provide information on the decay probability of the compound system. The apparently zero values of the distribution for $J>6$ reflects the fact that an angular momentum transfer of 4 is an upper classical limit for the given energy of the $^3$He projectile coming onto the target $^{157}$Gd. Notice that the formation probability P(J$^{\pi}$,E) resembles a Gaussian distribution but its magnitude and shape are different for positive and negative parity states.

\section{Concluding Remarks}
The present computational method is a first step to particle pick-up reactions in deformed  and  strongly deformed systems. 
Recent work on spherical systems has shown that pickup reactions which create highly-excited states, have important contributions from two-step processes, such as pickup followed by inelastic excitation and vice versa.  The present work needs to be extended to include those contributions.
%JE: There has been new theory development for stripping, so better steer clear of speculating on this. I can send you multiple papers, if you are interested.
%However, for stripping reactions the method would need to incorporate the effects of the pairing interaction since adding a nucleon would involve mostly levels near and above the Fermi level and even unbound states and resonances. In heavier nuclei the treatment of the valence single-particles states above the Fermi level would probably face a problem due to the treatment of weakly bound states that may render the current computational  technique useless; this could easily be monitored through evaluation of the norm of the bound states and the quality of the Sturmian basis. 

%JE: not needed if the next sentence is removed:
%The surrogate method (Fig. \ref{TheSurrogateMethod}) assumes that the surrogate reaction populates $J^{\pi}$ states of the  intermediate nuclear system within the same energy range as the desired reaction. 
%JE: this is not correct, I can tell you why in a longer email later.
%The method would certainly fail if the distribution $P(J^{\pi},E)$ becomes zero for some relevant values of $J^{\pi}$ at the relevant range of energies $E$ since there would be no way of deducing the decay probabilities $g_{\chi}(J^{\pi},E)$. If $P(J^{\pi},E)=0$ for a given $J^{\pi}$ at an isolated energy $E$ one could perhaps devise an analytic continuation for $g_{\chi}(J^{\pi},E)$. 
Our calculations show that within the assumptions outlined here, the one-step contribution to the reaction $^{3}$He+$^{157}$Gd $\rightarrow$ $^{4}$He+$^{156}$Gd$^{\star}$ has a smoothly-varying formation probability $P(J^{\pi},E)$ within a wide energy range relevant to the desired reaction $^{155}$Gd+n $\rightarrow$ $^{156}$Gd$^{\star}$. 
Thus, given an experimental input on the decay probability $P_{\delta\chi}$ into an exit channel $\chi$ within the surrogate formation channel $\delta$, one should in principle be able to determine $g_{\chi}(J^{\pi},E)$ using the approach outlined in Refs.~\cite{Escher:12rmp, Escher:16a}. 
%JE: Skip, as this may change with the inclusion of 2-step contributions:
%It is important to pint out that in our study the formation probability P(J$^{\pi}$,E) resembles a Gaussian distribution but its shape could be different for positive and negative parity states.

% JE: Not the best take-away point
%Experimental input is essential for the fine tuning of the model parameters such as single particle Woods-Saxon potential,  optical model potential for the incoming and outgoing projectiles as well as the choice of $\Gamma(E)$ used in the smearing function $\rho(E)$. Above all a comparison with experiment should be the true measure of the applicability the surrogate method.

\section*{Acknowledgments}
We thank Peter D. Kunz and Frank S. Dietrich for essential input and guidance and P. Navratil, J. P. Vary, and W. Younes for helpful discussions.
This work was partly performed under the auspices of the U. S. Department of Energy by the University of California, Lawrence Livermore National Laboratory under contract No. DE-AC52-07NA27344. 
%JE: Shortened acknowledgements and updated contract number. LDRD project too old to be mentioned here.

%\bibliographystyle{plane}
%\bibliographystyle{apsrev}
%\bibliographystyle{unsrt}
%\bibliographystyle{aipnum4-1}
%\bibliographystyle{iopart-num}
%\bibliographystyle{phaip}
%\bibliography{RCSforDN} \end{document}

\end{document}